\global\def\draftcontrol{0}

   \def\versionno{ Time-Dependent Processes -- draft   }

\catcode`\@=11

\expandafter\ifx\csname draftcontrol\endcsname\relax\global\def\draftcontrol{0}
\fi

{\count255=\time\divide\count255 by 60
\xdef\hourmin{\number\count255}
\multiply\count255 by-60\advance\count255 by\time
\xdef\hourmin{\hourmin:\ifnum\count255<10 0\fi\the\count255}}
\def\draftdate{\number\month/\number\day/\number\year\ \ \ \hourmin }

\newcommand\makepapertitle{\par
  \begingroup
    \renewcommand\thefootnote{\@fnsymbol\c@footnote}%
    \def\@makefnmark{\rlap{\@textsuperscript{\normalfont\@thefnmark}}}%
    \long\def\@makefntext##1{\parindent 1em\noindent
            \hb@xt@1.8em{%
                \hss\@textsuperscript{\normalfont\@thefnmark}}##1}%
     \newpage
     \global\@topnum\z@   
     \@makepapertitle
     \thispagestyle{empty}\@thanks
  \endgroup
  \setcounter{footnote}{0}%
  \global\let\thanks\relax
  \global\let\makepapertitle\relax
  \global\let\@makepapertitle\relax
  \global\let\@thanks\@empty
  \global\let\@author\@empty
  \global\let\@date\@empty
  \global\let\@title\@empty
  \global\let\title\relax
  \global\let\author\relax
  \global\let\date\relax
  \global\let\and\relax
  \def\version{\let\version\@version\@gobble}
}
\def\@makepapertitle{%
  \newpage
   \ifnum\draftcontrol=1 {}
   \version\versionno
   \vskip 3em%
   \else
   \hfill\hbox to 3cm {\parbox{4cm}{\@pubnum}\hss}%
   \vskip 3em%
   \fi
   \begin{center}%
   \let \footnote \thanks
     {\LARGE {\@title}}%
     \vskip 1.5em%
     {\normalsize
       \lineskip .5em%
       \begin{tabular}[t]{c}%
         \@author
       \end{tabular}\par}%
     \vskip 1.5em%
     {\@bstract}%
     \end{center}%
     \vskip 1.5em
     \@date%
   \par
}

\gdef\@pubnum{}
\def\pubnum#1{%
  \gdef\@pubnum{#1}}

\gdef\@bstract{}
\def\Abstract#1{%
  \gdef\@bstract{%
   \parbox{\textwidth-0pc}{%
   \centerline{\bf Abstract}\penalty1000%
\kern.2cm%
\noindent
\renewcommand\baselinestretch{1.0}%
{#1}}}
}

\def\ps@paper{\let\@mkboth\@gobbletwo%
     \ifnum\draftcontrol=1
    \def\@oddfoot{\hbox to \textwidth{\tiny \versionno \hfil\tiny\draftdate}%
    \hskip -\textwidth \hbox to \textwidth{\hfil\rm\thepage\hfil}}%
     \else\def\@oddfoot{\hbox to \textwidth{\hfil\rm\thepage\hfil}}
     \fi
     \let\@evenfoot\@oddfoot
}

\def\body{\clearpage
          \pagestyle{paper}
    }

\def\@version#1{\ifnum\draftcontrol=1
\typeout{}\typeout{#1}\typeout{}
\vskip3mm\centerline{\hbox{\fbox{\normalsize{\tt DRAFT -- #1 -- }
                   {\draftdate}}}}\vskip3mm
\fi}
\let\version\@version
\long\def\eqlabel#1{\ifnum\draftcontrol=1
                    \tag@false  
                    \tag*{(\theequation) \hbox to -0.2cm{\hspace{0cm}\small{#1}\hss}}
                    \refstepcounter{equation}
                    \edef\@currentlabel{\theequation}
                    \ltx@label{#1}          
                    \else
                    \label{#1}
                    \fi
                    }
\let\st@bibitem\@bibitem
\let\st@lbibitem\@lbibitem
\ifnum\draftcontrol=1
  \def\@bibitem#1{%
    \st@bibitem{#1}\a@@label{#1}\ignorespaces}
  \def\@lbibitem[#1]#2{%
    \st@lbibitem[#1]{#2}\a@@label{#2}\ignorespaces}
  \def\a@@label#1{%
    \gdef\a@lab{\smash{\normalfont\small#1}}
    \ifvmode
      \if@inlabel
        \global\setbox\@labels\hbox{%
          \llap{\a@lab\let\a@lab\relax
                \kern\@totalleftmargin\kern\marginparsep}%
          \box\@labels}%
      \fi
    \fi}
\fi

\documentclass[12pt,letterpaper]{article}

\usepackage{amsmath,amssymb,array,calc,epsfig}

\ifnum\draftcontrol=1
\fi

\renewcommand\baselinestretch{1.25}
\renewcommand\section{\@startsection {section}{1}{\z@}%
                                   {-3.5ex \@plus -1ex \@minus -.2ex}%
                                   {2.3ex \@plus.2ex}%
                                   {\normalfont\large\bfseries}}
\renewcommand\subsection{\@startsection{subsection}{2}{\z@}%
                                   {-3.25ex\@plus -1ex \@minus -.2ex}%
                                   {1.5ex \@plus .2ex}%
                                   {\normalfont\normalsize\bfseries}}
\renewcommand\subsubsection{\@startsection{subsubsection}{3}{\z@}%
                                   {-3.25ex\@plus -1ex \@minus -.2ex}%
                                   {1.5ex \@plus .2ex}%
                                   {\normalfont\normalsize\it}}
\renewcommand\paragraph{\@startsection{paragraph}{4}{\z@}%
                                   {-3.25ex\@plus -1ex \@minus -.2ex}%
                                   {1.5ex \@plus .2ex}%
                                   {\normalfont\normalsize\bf}}

\numberwithin{equation}{section}

\def\ie{{\it i.e.}}

\def\revise#1       {\raisebox{-0em}{\rule{3pt}{1em}}%
                     \marginpar{\raisebox{.5em}{\vrule width3pt\
                     \vrule width0pt height 0pt depth0.5em
                     \hbox to 0cm{\hspace{0cm}{%
                     \parbox[t]{4em}{\raggedright\footnotesize{#1}}}\hss}}}}

\newcommand\nxt[1]  {\\\fnxt#1}

\def\cali         {{\cal I}}

\def\calm         {{\cal M}}
\def\caln         {{\cal N}}
\def\calo         {{\cal O}}
\def\calp         {{\cal P}}

\def\calr         {{\cal R}}

\def\del          {\partial}

\def\tr           {\mathop{\rm Tr}}

\def\sqr#1#2{{\vcenter{\vbox{\hrule height.#2pt
 \hbox{\vrule width.#2pt height#1pt \kern#1pt
 \vrule width.#2pt}\hrule height.#2pt}}}}

\newcommand{\ft}[2]{{\textstyle{\frac{#1}{#2}}}}

\def\a{\alpha}
\def\b{\beta}

\def\r{\rho}

\def\e{\epsilon}

\def\aa1{\phi}
\def\cc1{\psi}

\def\t{\tau}

\def\om{\Omega}
\def\l{\lambda}

\newcommand{\eq}{\begin{equation}}
\newcommand{\eqx}{\end{equation}}
\newcommand{\eqn}{\begin{eqnarray}}
\newcommand{\eqnx}{\end{eqnarray}}
\newcommand{\f}[2]{\frac{#1}{#2}}

\catcode`\@=12

\begin{document}


\title{\bf On the supergravity description of boost invariant conformal plasma at strong coupling}
\pubnum{%
UWO-TH-07/18}

\date{December 2007}

\author{
Paolo Benincasa$ ^1$,  Alex Buchel$ ^{1,2}$, Michal P. Heller$ ^3$ and Romuald A. Janik$ ^3$\\[0.4cm]
\it $ ^1$Department of Applied Mathematics\\
\it University of Western Ontario\\
\it London, Ontario N6A 5B7, Canada\\[0.2cm]
\it $ ^2$Perimeter Institute for Theoretical Physics\\
\it Waterloo, Ontario N2J 2W9, Canada\\
\it $ ^3$ Institute of Physics\\
\it Jagellonian University\\
\it Reymonta 4, 30-059 Krakow, Poland
}

\Abstract{We study string theory duals of the expanding boost invariant
conformal gauge theory plasmas at strong coupling. The dual
supergravity background is constructed as an asymptotic late-time
expansion, corresponding to equilibration of the gauge theory
plasma. The absence of curvature singularities in the first few orders
of the late-time expansion of the dual gravitational background
unambiguously determines the equilibrium equation of the state, and
the shear viscosity of the gauge theory plasma. While the absence of
the leading pole singularities in the gravitational curvature
invariants at the third order in late-time expansion determines the
relaxation time of the plasma, the subleading logarithmic singularity
can not be canceled within a supergravity approximation.  Thus, a
supergravity approximation to a dual description of the strongly
coupled boost invariant expanding plasma is inconsistent. Nevertheless
we find that the relaxation time determined from cancellation of pole
singularities is quite robust.
}

\makepapertitle

\body

\version\versionno

\section{Introduction}

Gauge/string correspondence \cite{m1,m2} developed into a valuable tool
to study strong coupling dynamics of gauge theory plasma, and might be relevant in understanding the
physics of quark-gluon plasma (QGP) produced at RHIC in heavy ion collisions \cite{rhic1,rhic2,rhic3,rhic4}.
Besides applications devoted to determine the equation of state of the gauge theory plasma at strong coupling
\cite{ads1,ads2,nc1,nc2,nc3}, the dual string theory models have been successful in computing plasma transport
properties \cite{u1,u2,u3,bulk,cor1,cor2,mt}, the quenching of partonic jets \cite{lrw,hkkky,aev}, and
photon and dilepton production \cite{CaronHuot:2006te}. Much less studied are the truly dynamical processes in
plasma\footnote{One notable example is the analysis of the shock waves generated by a heavy quark moving in gauge theory
plasma \cite{sw1,sw2}.} .

In \cite{j1} a framework of constructing the string theory duals to
expanding boost invariant conformal plasma
was proposed\footnote{Earlier work on more qualitative aspects of
expanding plasma includes \cite{zahed}.}. This was later used to study
viscous properties of the expanding plasma \cite{shin1,j2,j3}, other
applications were further discussed in
\cite{siopsis,ext0,shin2,kaj1,ext1,kov,kaj2,ext2}.  
The basic idea of the approach is to set up the asymptotic geometry of the gravitational dual to that of the boost invariant
frame, suggested by Bjorken \cite{bj} as a description of the central rapidity region in highly relativistic nucleus-nucleus
collisions. One further has to specify the normalizable modes (a string theory duals to expectation values of
appropriate gauge invariant operators in plasma) for the metric and other supergravity/string theory fields so that
the singularity-free description of string theory is guaranteed. As explained in \cite{j1,j2,j3}, requiring the absence of
certain curvature singularities in dual gravitational backgrounds constrains the equilibrium equation of state of supersymmetric $\caln=4$
Yang-Mills (SYM) plasma (in agreement with \cite{ads1}), its shear viscosity (in agreement with \cite{pss}), and the relaxation time
of $\caln=4$ SYM plasma (in agreement with \cite{st}).  Despite an apparent success of the framework, it suffers a serious drawback:
not all the singularities in the dual string theory description of the strongly coupled expanding $\caln=4$ SYM plasma have been canceled
\cite{j3}. The subject of this paper is to comment on the gravitational singularities observed in \cite{j3}.

To begin with, we would like to emphasize the importance of singularities in string theory. As we already mentioned,
constructing a string theory dual to expanding boost-invariant plasma implied identification of a set of string theory
fields with non-vanishing normalizable modes, dual to vacuum expectation values (VEVs) of the gauge invariant operators in plasma.
A priori, it is difficult to determine which operators in plasma will develop a vacuum expectation value \footnote{For
a weak coupling analysis in QGP pointing to a development of a nonzero expectation value for
$\tr F^2$ see \cite{rrs}.}. In practice, on the dual gravitational side of the correspondence one usually truncates
the full string theory to low-energy type IIB supergravity, and often further to a consistent Kaluza-Klein
truncation of type IIB supergravity, keeping supergravity modes invariant under the symmetries of the problem.
From the gauge theory perspective, such an approximation implies that only certain operators
are assumed to develop a VEV. Typically these operators are gauge-invariant operators of
low dimension. Of course, the latter assumption might, or might not be correct:
luckily, on the gravitational side the consistency of the approximation is severely constrained by requiring an
absence of singularities in the bulk of the gravitational background. A typical example of this phenomena is
thermodynamics of cascading gauge theory \cite{kl}: if one assumes that the only operators that develop a VEV
at high temperatures in this gauge theory are the same ones as those at zero temperature, all dual gravitational backgrounds
will have a null singularity \cite{kt1,kt2}; on the other hand, turning on a VEV of an irrelevant  dimension-6 operator
\cite{aby1} leads to a smooth geometry \cite{kt3}. Thus, the presence of bulk curvature singularities in
gravitational backgrounds of string theory points to an inconsistent truncation.

In \cite{j1,j2,j3} the authors assumed a truncation of the full string theory dual to $\caln=4$ expanding boost-invariant SYM plasma to
a Kaluza-Klein reduction of type IIB supergravity on $S^5$, maintaining only the normalizable modes of the five-dimensional metric
(dual to a stress-energy tensor of the plasma), and the dilaton normalizable mode (dual to a dimension-4 operator $\tr F^2$).
The absence of  singularities in five-dimensional Riemann tensor invariant $R_{\mu\nu\l\r}^{(5)}R^{(5)\mu\nu\l\r}$ of the asymptotic
late-time expansion
of the background geometry at leading and first two subleading orders correctly reproduced the equilibrium equation of state of $\caln=4$
SYM plasma and its shear viscosity. Furthermore, removing the pole singularities in $R_{\mu\nu\l\r}^{(5)}R^{(5)\mu\nu\l\r}$
at third order in late-time expansion
determined for the first time the relaxation time of the strongly coupled $\caln=4$ SYM plasma \cite{j3}. Given the truncation of the
string theory, it was not possible to cancel a logarithmic singularity at third order in $R_{\mu\nu\l\r}^{(5)}R^{(5)\mu\nu\l\r}$.
Turning on an appropriate  dilaton mode removes a logarithmic singularity
in the five-dimensional string frame metric in
$R_{\mu\nu\l\r}^{(5,string)}R^{(5,string)\mu\nu\l\r}$. However, as we
show in the present paper, the full 10 dimensional metric remains
singular even in the string frame.

The paper is organized as follows. In the next section we review the boost invariant kinematics of gauge theory plasma.
We identify parameters in the late-time expansion of the stress-energy tensor one-point correlation function with the hydrodynamic
parameters of M{\"u}ller-Israel-Stewart transient theory \cite{m,is}.
Following the idea that a presence of singularities indicates an inconsistent truncation, we consider in the section 3 the full  $SO(6)$ invariant
sector of type IIB supergravity, assuming the parity invariance of the
$\caln=4$ SYM expanding plasma. Compare to the analysis
in \cite{j3}, this includes one additional $SO(6)$ invariant scalar, dual to a dimension-8 operator. We show that using this additional
scalar one can cancel the logarithmic singularity at the third order
in late-time expansion either in ten-dimensional
Einstein frame Ricci tensor squared $\calr_{\mu\nu}\calr^{\mu\nu}$, or
in ten-dimensional
Einstein frame Riemann tensor squared
$\calr_{\mu\nu\l\r}\calr^{\mu\nu\l\r}$, but not {\it both} the
singularities.
Given the {\em assumed} symmetries of the expanding plasma system,
there are no additional supergravity fields to be turned on to cancel
the singularity in string theory dual to
expanding $\caln=4$ SYM plasma. In section 4 we further show
that canceling a singularity
typically requires additional supergravity (or string theory) fields
to be turned on. Using the consistent truncation
presented in \cite{aby1,bkt}, we consider supergravity dual to
superconformal Klebanov-Witten gauge theory \cite{kw} plasma.
Here, in addition to a dimension-8 operator (analogous to the one
discussed in section 3) there is an additional scalar, dual to
a dimension-6 operator \cite{aby1}. We show that turning on the latter scalar can remove logarithmic singularities in the third order
in the late-time expansion of the  ten-dimensional
Einstein frame Ricci scalar $\calr$, as well as
$\calr_{\mu\nu}\calr^{\mu\nu}$ and
$\calr_{\mu\nu\l\r}\calr^{\mu\nu\l\r}$.
However, we find new logarithmic singularities at the third order in higher curvature
invariants such as $\calr_{\mu_1\nu_1\l_1\r_1} \calr^{\mu_1\nu_1\l_2\r_2} \calr_{\mu_2\nu_2}\ ^{\l_1\r_1}\calr^{\mu_2\nu_2}\ _{\l_2\r_2}$,
as well as  logarithmic
singularities with  different coefficients in $(\calr_{\cdot\cdot\cdot\cdot})^8$, $(\calr_{\cdot\cdot\cdot\cdot})^{16}$ and so on.
Canceling  the logarithmic singularities in all the curvature invariants appears to require an infinite number of fields.
Since such fields are absent in supergravity approximation to string theory dual of expanding boost invariant
plasma, we conclude that such a supergravity approximation is inconsistent. In section 5 we present further directions in
analyzing the time-dependent framework proposed in \cite{j1,j2,j3}.

\section{Review of boost-invariant kinematics}

Quark-gluon plasma is expected to be produced in high-energy collisions of heavy
ions. In nucleus-nucleus experiments a central rapidity plateau structure for
particle formation has been observed. Specifically, the expansion of the plasma
appears to be longitudinal and homogeneous near the collision axis in the case of
central collisions. Thus the system is expected to be boost-invariant in the
the longitudinal plane. It is convenient to introduce
the proper-time $\tau$ and the rapidity $y$, defined as
\eq
x^0=\tau \cosh y\,,  \quad\quad\quad\quad x^3=\tau \sinh y\,.
\eqx
Then rapidity invariance amounts to independence on $y$. We will also
assume $y \to -y$ invariance.

We will also make another approximation, following \cite{bj},
namely independence on the transverse coordinates $x_\perp$. This
corresponds to the limit of very large nuclei.

With these assumptions, the whole energy-momentum tensor can be
expressed purely in terms of a single function\footnote{After taking
into account energy-momentum conservation and tracelessness. For
explicit formulas see \cite{j1}.}
$\e(\tau)$, the energy density in the local rest frame. This remaining
function should be fixed by gauge theory dynamics.

The proposal of \cite{j1} to use the AdS/CFT correspondence to
determine $\e(\tau)$ amounts to constructing a dual geometry, by
solving supergravity equations with the boundary conditions
\eq
\left\langle T_{\mu\nu} \right\rangle =\f{N_c^2}{2\pi} \lim_{z\to 0}
\f{g_{\mu\nu}-\eta_{\mu\nu}}{z^4}\,,
\eqx
where the (5-dimensional part of the) metric is written in the
Fefferman-Graham form:
\eq
ds_{5dim.}^2=\f{g_{\mu\nu} dx^\mu dx^\nu+dz^2}{z^2}\,.
\eqx

Given any $\e(\tau)$ it is possible to construct such a metric. In
\cite{j1} it was advocated that the requirement of the non-singularity of this
geometry picks out a physically acceptable $\e(\tau)$. In \cite{j1,j2,j3}
this problem was analyzed in an expansion for large $\tau$. At each
order the square of the Riemann curvature tensor had poles which could
be canceled only by a specific choice of the coefficients of
$\e(\tau)$. In the third coefficient there remained a leftover
logarithmic singularity in the Einstein frame\footnote{ The is also a singularity
in the string frame Ricci tensor squared $R_{\mu\nu}^{(10,string)}R^{(10,string)\mu\nu}$.} . In this paper we
would like to revisit this issue.

The outcome of this calculation is a specific $\tau$ dependence for
$\e(\tau)$:
\eq
\label{etau}
\e(\tau)=\f{1}{\tau^{\f{4}{3}}}\left(1- \f{\ldots}{\tau^{\f{2}{3}}}+
\f{\ldots}{\tau^{\f{4}{3}}} +\ldots \right)\,.
\eqx

Now we may obtain the physical interpretation of the above result by
checking whether it is a solution of some specific phenomenological
equations. In the above case, the
leading behavior corresponds to perfect fluid
hydrodynamics, the subleading one is related to the effect of shear
viscosity, while the third one is related to second order viscous
hydrodynamics relaxation time. Let us note that the relaxation
time depends on the form of phenomenological equations used to
interpret (\ref{etau}).

M{\"u}ller-Israel-Stewart theory \cite{m,is} of dissipative processes provides a nice framework to study relativistic fluid dynamics.
In the Bjorken regime of boost invariant expansion of plasma, the transport equations take the form \cite{mur}
\begin{equation}
\begin{split}
0=&\frac{d\e}{d\t}+\frac{\e+p}{\t}-\frac 1\t \Phi\\
0=&\frac{d\Phi}{d\t}+\frac{\Phi}{\t_\pi}+\frac 12 \Phi\left(\frac 1\t +\frac{1}{\b_2}T\frac{d}{d\t}\left(\frac{\b_2}{T}\right)\right)
-\frac 23 \frac{1}{\b_2}\frac 1\t\,,
\end{split}
\eqlabel{transport}
\end{equation}
where $\t_\pi$ is the relaxation time,
$\e$ is the energy density, $p$ is the pressure, $\Phi$ is related to the dissipative part of the energy-momentum,
and
\begin{equation}
\b_2=\frac{\t_\pi}{2\eta}\,,
\eqlabel{b2def}
\end{equation}
with $\eta$ being the shear viscosity.
In \eqref{transport} anticipating application to conformal gauge
theory plasma we set the bulk viscosity $\zeta$ to zero.

In the case of the $\caln=4$ SYM plasma we have
\begin{equation}
\begin{split}
&\e(\t)=\frac{3}{8}\pi^2 N^2\ T(\t)^4\,,\qquad p(\t)=\frac 13\e(\t)\,,\qquad \eta(\t)=A\ s(\t)=A\  \frac{1}{2}\pi^2 N^2\ T(\t)^3\,,\\
&\t_\pi(\t)=r\ \t_\pi^{\rm Boltzmann}(\t)=r\ \frac{3\eta(\t)}{2p(\t)}\,.
\end{split}
\eqlabel{cftp}
\end{equation}
Given \eqref{cftp}, we can solve \eqref{transport} perturbatively as $\t\to\infty$:
\begin{equation}
T(\t)=\frac{\Lambda}{\t^{1/3}}\ \left(1+\sum_{k=1}^\infty \frac{t_k}{(\Lambda\t^{2/3})^k}\right)\,,\qquad \Phi(\t)=\frac 23\pi^2 N^2\ A\
\frac{\Lambda^3}{\t^2}\ \left(1+\sum_{k=1}^\infty \frac{f_k}{(\Lambda\t^{2/3})^k}\right)\,,
\eqlabel{solpert}
\end{equation}
where $\Lambda$ is an arbitrary scale. The constant coefficients $\{t_k,f_k\}$ are related to the $\caln=4$ SYM dimensionless
transport parameters $A$ ( the ratio of shear-to-entropy density  ) and $r$ ( the  plasma relaxation time in units
of Boltzmann relaxation time ). For the first few coefficients we find:
\begin{equation}
\begin{split}
&t_1=-\frac 23 A\,,\qquad t_2=-\frac 43 A^2 r\,,\qquad t_3=-\frac{8}{27}A^3 r(1+24r)\,,\\
&f_1=2A(2r-1)\,,\qquad f_2=\frac 43 A^2(1-8r+24r^2)\,,\\
 &f_3=\frac{8}{27}A^3(-1+24 r-234r^2+1296r^3)\,.
\end{split}
\eqlabel{coeff}
\end{equation}

From the supergravity computations first done in  \cite{j3} (and reproduced in the following section of  this paper) we find
\begin{equation}
\e(\t)=\left(\frac{N^2}{2\pi^2}\right)\ \frac{1}{\t^{4/3}}\biggl\{1-\frac{2\eta_0}{\t^{2/3}}+\left(\frac{10}{3}\eta_0^2+\frac{C}{36}
\right)\frac{1}{\t^{4/3}}+\cdots\biggr\}\,.
\eqlabel{egrav}
\end{equation}
Matching the gauge theory expansion for the energy density \eqref{cftp} with that of the dual gravitational description \eqref{egrav}
we find
\begin{equation}
\Lambda=\frac{\sqrt{2}}{3^{1/4}\pi }\,,\qquad A=\frac{3^{3/4}}{2^{3/2}\pi}\ \eta_0\,,
\qquad r=-\frac{11}{18}-\frac{1}{108}\ \frac{C}{\eta_0^2}\,.
\eqlabel{ma1}
\end{equation}
Using the supergravity results \eqref{eta0} and \eqref{cdef}, we find
\begin{equation}
A=\frac{1}{4\pi}\,,\qquad r=\frac 13(1-\ln 2)\,.
\eqlabel{ma2}
\end{equation}

A different formulation of second order hydrodynamics in \cite{BRW}
derived from Boltzmann equations lead to the equation
\begin{equation}
\begin{split}
0=&\frac{d\e}{d\t}+\frac{\e+p}{\t}-\frac 1\t \Phi\\
0=&\frac{d\Phi}{d\t}+\frac{\Phi}{\t^{(ii)}_\pi} - \frac 23
\frac{1}{\b_2}\frac 1\t\,,
\end{split}
\eqlabel{transport2}
\end{equation}
which provides a different definition of the relaxation time
$\t^{(ii)}_\pi$. Using (\ref{transport2}) leads to $r^{(ii)}=\frac 19(1-\ln
2)$ quoted in \cite{j3}. Of course (\ref{transport}) and
(\ref{transport2}) provide different interpretations of the same $\e(\tau)$.
Without lifting the symmetry assumptions of the uniform
boost invariant plasma expansion we may not rule out one of the
(\ref{transport}) and (\ref{transport2}) descriptions.

\section{$\caln=4$ QGP}
Supergravity dual to boost invariant expanding $\caln=4$ QGP was discussed in \cite{j1,j2,j3}. We extend the previous analysis
including {\it all} supergravity modes, invariant under the symmetries of the problem. Specifically,
we assume that the $SO(6)$ R-symmetry of $\caln=4$ SYM is unbroken. We further assume that the boost invariant plasma is
invariant under separate reflections of all three spatial directions. The latter assumption
in particular ensures that the supergravity dual excludes the axion and various 3-form fluxes along extended spacial
directions of the boundary --- say RR fluxes with components  $F_{\t x_1x_2}$ within the metric ansatz \eqref{5dimM}.
In section 5 we comment why we believe that even allowing for a parity violating supergravity
modes would not change our main conclusion, \ie, the inconsistency of the supergravity approximation
in dual description of boost invariants expanding plasmas.

Given the symmetry assumptions of the problem as well as the approximation of the full string theory by its low energy
type IIB supergravity, the non-vanishing fields we need to keep is the five-dimensional metric, the warp factor of the
five-sphere, the self-dual five-form flux and the dilaton. Parity invariance and the symmetries of the
boost invariant frame further restricts the five-dimensional metric as discussed below.

\subsection{Consistent Kaluza-Klein Reduction}

Consider Einstein frame type IIB low-energy effective action in $10$-dimensions
\begin{equation}\eqlabel{10dimS}
 S_{10}\:=\:\frac{1}{2\kappa_{10}^{2}}
  \int_{\mathcal{M}_{10}}d^{10}\xi\:\sqrt{-\tilde{g}}
  \left\{
   \calr
   -\frac{1}{2}\left(\tilde{\partial}\phi\right)^{2}
   -\frac{1}{4\cdot 5!}F_{5}^{2}
  \right\},
\end{equation}
with metric ansatz:
\begin{equation}\eqlabel{10dimM}
 \begin{split}
  d\tilde{s}_{10}^{2}\:&=\:\tilde{g}_{MN}d\xi^{M}d\xi^{N}\:=\\
  &=\:\sigma^{-2}(x)g_{\mu\nu}(x)dx^{\mu}dx^{\nu}+
      \sigma^{6/5}(x)\left(dS^{5}\right)^{2}\,,
 \end{split}
\end{equation}
where $M,\,N,\,\ldots\,=\,0,\ldots,9$ and $\mu,\,\nu,\,\ldots\,=\,0,\ldots,4$,
and $(dS^{5})^{2}$ is the line element for a $5$-dimensional sphere with unit
radius. For the $5$-form $F_{5}$ we assume that
\begin{equation}\eqlabel{5form}
F_{5}\:=\:\mathcal{F}_{5}+\star\mathcal{F}_{5}\,,\qquad
\mathcal{F}_{5}\:=\:-4 Q\ \omega_{S^{5}}\,,
\end{equation}
where $\omega_{S^{5}}$ is the 5-sphere volume form and $Q$ is a constant.
We further assume that the dilaton is $\phi=\phi(x)$.

With the ansatz \eqref{10dimM}, we find
\begin{equation}\eqlabel{red}
 \begin{split}
  &\sqrt{-\tilde{g}}\:=\:\sigma^{-2}\sqrt{-g}\sqrt{g_{(\mbox{\tiny $S^{5}$})}}\\
  &\calr\:=\:
   \sigma^{2}\left\{R-\frac{24}{5}\left(\partial\ln{\sigma}\right)^{2}\right\}+
   20\sigma^{-6/5}\\
  &-\frac{1}{4\cdot 5!}F_{5}^{2}\:=\:-8 Q^2\sigma^{-6}\\
&  -\frac{1}{2}\left(\tilde{\partial}\phi\right)^{2}=  -\frac{1}{2}\left(\partial\phi\right)^{2}\ \sigma^2\,,
 \end{split}
\end{equation}
where $R$ is the Ricci scalar for the $5$-dimensional metric $g_{\mu\nu}$.
The $5$-dimensional effective action therefore takes the following form:
\begin{equation}\eqlabel{5dimS}
 S_{5}^{\mbox{\tiny eff}}\:=\:
  \frac{1}{2\kappa_{5}^{2}}\int_{\mathcal{M}_{5}}d^{5}x\:\sqrt{-g}
  \left\{
   R -\frac{1}{2}\left(\partial\phi\right)^{2}-\frac{24}{5}\left(\partial\ln{\sigma}\right)^{2}-
   \mathcal{P}(\sigma)
  \right\}\,,
\end{equation}
with
\begin{equation}\eqlabel{pot}
\mathcal{P}(\sigma)\:=\:-20\sigma^{-16/5}+8 Q^2\sigma^{-8}\,,\qquad
\kappa_{5}^{2}\:=\:\frac{\kappa_{10}^{2}}{\mbox{\footnotesize vol}
  \left\{S^{5}\right\}}\,.
\end{equation}

It is convenient to introduce a scalar field $\alpha(x)$ defined as
\begin{equation}\eqlabel{alpha}
\sigma(x)\:=\:e^{\alpha(x)}\,,
\end{equation}
so that \eqref{5dimS} can be rewritten as:
\begin{equation}\eqlabel{5dimS2}
 S_{5}^{\mbox{\tiny eff}}\:=\:
  \frac{1}{2\kappa_{5}^{2}}\int_{\mathcal{M}_{5}}d^{5}x\:\sqrt{-g}
  \left\{
   R -\frac{1}{2}\left(\partial\phi\right)^{2}-\frac{24}{5}\left(\partial\alpha\right)^{2}-
   \mathcal{P}(\alpha)
  \right\}\,,
\end{equation}
where
\begin{equation}\eqlabel{pota}
 \mathcal{P}(\alpha)\:=\:-20\,e^{-16\alpha/5}+8 Q^2\,e^{-8\alpha}\,.
\end{equation}
From \eqref{5dimS2}, the Einstein equations and the equation of motion for
$\alpha$ are respectively given by:
\begin{equation}\eqlabel{eom}
 \begin{split}
  &R_{\mu\nu}\:=\:
   \frac{24}{5}\left(\partial_{\mu}\alpha\right)\left(\partial_{\nu}\alpha\right)+  \frac{1}{2}\left(\partial_{\mu}\phi\right)
   \left(\partial_{\nu}\phi\right)+\frac{1}{3}g_{\mu\nu}\mathcal{P}(\alpha)\\
  &\Box\alpha\:=\:\frac{5}{48}\frac{\partial\mathcal{P}}{\partial\alpha}\\
&\Box\phi\:=\:0\,.
 \end{split}
\end{equation}

\subsection{Equations of motion}

For the five-dimensional metric we use the same ansatz as in \cite{j1,j2,j3}:
\begin{equation}\eqlabel{5dimM}
 \begin{split}
  ds^{2}\:&=\:g_{\mu\nu}dx^{\mu}dx^{\nu}\:=\:\\
          &=\frac{1}{z^{2}}
	    \left[
             -e^{2a(\tau,z)}d\tau^{2}+e^{2b(\tau,z)}\tau^{2}dy^{2}+
	      e^{2c(\tau,z)}dx_{\perp}^{2}
	    \right]+
	    \frac{dz^{2}}{z^{2}}\,,
 \end{split}
\end{equation}
where $dx_{\perp}^{2}\equiv dx_1^2+dx_2^2$.
It is easy to see that this is the most general boost invariant geometry
written in Fefferman-Graham coordinates subject to a parity
invariance along the boost direction\footnote{The parity invariance
excludes metric components $g_{yz}(\t,z)$ and $g_{\t y}(\t,z)$, in principle
allowed by a Fefferman-Graham coordinate frame. }.

Further assuming  $\alpha=\alpha(\tau,z)$ and $\phi=\phi(\tau,z)$, the equations of motion
for the background \eqref{5dimM}, the dilaton $\phi$ and the scalar $\alpha$
become:
\begin{itemize}
\item Einstein equations
\begin{equation}\eqlabel{Rtt}
 \begin{split}
  &e^{2a}
  \left[
    \partial_{z}^{2}a+\left(\partial_{z}a\right)^{2}+
    \left(\partial_{z}a\right)\left(\partial_{z}b\right)+
    2\left(\partial_{z}a\right)\left(\partial_{z}c\right)-
    \frac{4}{z}\partial_{z}a-\frac{1}{z}\partial_{z}b-\frac{2}{z}\partial_{z}c+
    \frac{4}{z^{2}}
   \right]\\
  & -
  \left[
  \partial_{\tau}^{2}b+\left(\partial_{\tau}b\right)^{2}-
    \left(\partial_{\tau}a\right)\left(\partial_{\tau}b\right)+
    2\left(\partial_{\tau}^{2}c-\left(\partial_{\tau}c\right)^{2}-
    \left(\partial_{\tau}a\right)\left(\partial_{\tau}c\right)
   \right)-
   \frac{1}{\tau}\partial_{\tau}a+\frac{2}{\tau}\partial_{\tau}b
  \right]  \\
  & = \frac{24}{5}\left(\partial_{\tau}\alpha\right)^{2}+
      \frac{1}{2}\left(\partial_{\tau}\phi\right)^{2}-
      \frac{1}{3}\frac{e^{2a}}{z^{2}}\mathcal{P}(\alpha)\,,
 \end{split}
\end{equation}
\begin{equation}\eqlabel{Ryy}
 \begin{split}
  &e^{2a}
  \left[
   \partial_{z}^{2}b+\left(\partial_{z}b\right)^{2}+
   \left(\partial_{z}a\right)\left(\partial_{z}b\right)+
   2\left(\partial_{z}b\right)\left(\partial_{z}c\right)-
   \frac{1}{z}\partial_{z}a-\frac{4}{z}\partial_{z}b-\frac{2}{z}\partial_{z}c+
   \frac{4}{z^{2}}
  \right]\\
  &-
  \left[
   \partial_{\tau}^{2}b+\left(\partial_{\tau}b\right)^{2}-
   \left(\partial_{\tau}a\right)\left(\partial_{\tau}b\right)+
   2\left(\partial_{\tau}b\right)\left(\partial_{\tau}c\right)-
   \frac{1}{\tau}\partial_{\tau}a+
   \frac{2}{\tau}\partial_{\tau}b+\frac{2}{\tau}\partial_{\tau}c
  \right]\\
  &=-\frac{1}{3}\frac{e^{2a}}{z^{2}}\mathcal{P}(\alpha)\,,
 \end{split}
\end{equation}
\begin{equation}\eqlabel{Rxx}
 \begin{split}
  &e^{2a}
  \left[
   \partial_{z}^{2}c+2\left(\partial_{z}c\right)^{2}+
   \left(\partial_{z}a\right)\left(\partial_{z}c\right)+
   \left(\partial_{z}b\right)\left(\partial_{z}c\right)-
   \frac{1}{z}\partial_{z}a-\frac{1}{z}\partial_{z}b-\frac{5}{z}\partial_{z}c+
   \frac{4}{z^{2}}
  \right]\\
  &-
  \left[
   \partial_{\tau}^{2}c+2\left(\partial_{\tau}c\right)^{2}-
   \left(\partial_{\tau}a\right)\left(\partial_{\tau}c\right)+
   \left(\partial_{\tau}b\right)\left(\partial_{\tau}c\right)+
   \frac{1}{\tau}\left(\partial_{\tau}c\right)
  \right]  \\
  & =
  -\frac{1}{3}\frac{e^{2a}}{z^{2}}\mathcal{P}(\alpha)\,,
 \end{split}
\end{equation}
\begin{equation}\eqlabel{Rzz}
 \begin{split}
   &\partial_{z}^{2}a+\left(\partial_{z}a\right)^{2}+
   \partial_{z}^{2}b+\left(\partial_{z}b\right)^{2}+
   2\left[\partial_{z}^{2}c+\left(\partial_{z}c\right)^{2}\right]-
   \frac{1}{z}\partial_{z}a-\frac{1}{z}\partial_{z}b-\frac{2}{z}\partial_{z}c+
   \frac{4}{z^{2}}\\
   &=
   -\frac{24}{5}\left(\partial_{z}\alpha\right)^{2}-
    \frac{1}{2}\left(\partial_{z}\phi\right)^{2}-
    \frac{1}{3}\frac{\mathcal{P}(\alpha)}{z^{2}}\,,
 \end{split}
\end{equation}
\begin{equation}\eqlabel{Rtz}
 \begin{split}
  &\partial_{\tau}\partial_{z}b-
  \left(\partial_{\tau}b\right)\left(\partial_{z}a\right)+
  \left(\partial_{\tau}b\right)\left(\partial_{z}b\right)+
  2\left[
   \partial_{\tau}\partial_{z}c+
   \left(\partial_{\tau}c\right)\left(\partial_{z}c\right)-
   \left(\partial_{\tau}c\right)\left(\partial_{z}a\right)
  \right]\\
&-
  \frac{1}{\tau}\partial_{z}a+\frac{1}{\tau}\partial_{z}b
  =
  -\frac{24}{5}
   \left(\partial_{\tau}\alpha\right)\left(\partial_{z}\alpha\right)-
   \frac{1}{2}\left(\partial_{\tau}\phi\right)\left(\partial_{z}\phi\right)\,;
 \end{split}
\end{equation}

\item dilaton equation:
\begin{equation}\eqlabel{dilaton}
 \begin{split}
  &0 =z^{2}
    \left\{
     \partial_{z}^{2}\phi+
     \left(\partial_{z}a+\partial_{z}b+2\partial_{z}c\right)
     \left(\partial_{z}\phi\right)-\frac{3}{z}\left(\partial_{z}\phi\right)
    \right.\\
    &\left.
     -e^{-2a}
     \left[
      \partial_{\tau}^{2}\phi +
      \left(
       -\partial_{\tau}a + \partial_{\tau}b + 2\partial_{\tau}c+\frac{1}{\tau}
      \right)
      \left(\partial_{\tau}\phi\right)
     \right]
    \right\}\,;
 \end{split}
\end{equation}

\item equation of motion for $\alpha$
\begin{equation}\eqlabel{alphaeq}
 \begin{split}
   \frac{5}{48}\frac{\partial\mathcal{P}}{\partial\alpha} &=z^{2}
       \left\{
        \partial_{z}^{2}\alpha+
        \left(\partial_{z}a+\partial_{z}b+2\partial_{z}c\right)
        \left(\partial_{z}\alpha\right)-
	\frac{3}{z}\left(\partial_{z}\alpha\right)
       \right.\\
     &\left.
      -e^{-2a}
       \left[
        \partial_{\tau}^{2}\alpha +
        \left(
        -\partial_{\tau}a + \partial_{\tau}b + 2\partial_{\tau}c+\frac{1}{\tau}
        \right)
        \left(\partial_{\tau}\alpha\right)
       \right]
      \right\}\,.
 \end{split}
\end{equation}

\end{itemize}

\subsection{Late-time expansion}
In \cite{j1,shin1,j2,j3} equations similar to \eqref{Rtt}-\eqref{dilaton} were solved asymptotically as a  late-time
expansion in powers of $(\t^{-2/3})$  (or powers of $(\t^{-1/3})$ for the dilaton) introducing the
scaling variable
\begin{equation}
v\equiv \frac{z}{\t^{1/3}}
\eqlabel{vdef}
\end{equation}
in the limit $\t\to \infty$, with $v$ kept fixed. Such a scaling is well motivated on physical grounds as in this case
we find that the boundary energy density $\e(\t)$ extracted from the one-point correction function of the
boundary stress-energy tensor
\begin{equation}
\e(\t)=-\frac{N^2}{2\pi^2}\lim_{z\to 0}\ \frac{2 a(z,\t)}{z^4}=
       -\frac{N^2}{2\pi^2}\lim_{v\to 0}\ \frac{2 a(v,\t)}{v^4\t^{4/3}}
\eqlabel{et}
\end{equation}
would have a late-$\t$ expansion appropriate for a conformal plasma in Bjorken regime \cite{j3}.
Thus  we  expect
\begin{equation}
\begin{split}
&a(\tau,v) = a_{0}(v) + \frac{1}{\tau^{2/3}} a_{1}(v) + \frac{1}{\tau^{4/3}} a_{2}(v) + \frac{1}{\tau^{2}} a_{3}(v)+\calo(\t^{-8/3})\\
&b(\tau,v) = b_{0}(v) + \frac{1}{\tau^{2/3}} b_{1}(v) + \frac{1}{\tau^{4/3}} b_{2}(v) + \frac{1}{\tau^{2}} b_{3}(v)+\calo(\t^{-8/3})\\
&c(\tau,v) = c_{0}(v) + \frac{1}{\tau^{2/3}} c_{1}(v) + \frac{1}{\tau^{4/3}} c_{2}(v) + \frac{1}{\tau^{2}} c_{3}(v)+\calo(\t^{-8/3})\,.\\
\end{split}
\eqlabel{ltseries1}
\end{equation}
In the equilibrium, both the dilaton and the $\a$-scalar vanish, implying that in \eqref{Rtt}-\eqref{Rtz} they enter quadratically
for small $\phi$, $\a$. Thus, given \eqref{ltseries1} we have two possible late-time asymptotic expansions for $\phi$ and $\a$:
\begin{equation}
\begin{split}
&\phi(\tau,v) =\frac{1}{\tau^{2/3}} \phi_{1}(v) + \frac{1}{\tau^{4/3}} \phi_{2}(v) + \frac{1}{\tau^{2}} \phi_{3}(v)+\calo(\t^{-8/3})\\
&{\rm or}\\
&\phi(\tau,v) =\frac{1}{\tau^{1/3}} \hat{\phi}_{1}(v) + \frac{1}{\tau} \hat{\phi}_{2}(v) + \frac{1}{\tau^{5/3}} \hat{\phi}_{3}(v)
+\calo(\t^{-7/3})\,,
\end{split}
\eqlabel{ltseries2}
\end{equation}
and
\begin{equation}
\begin{split}
&\a(\tau,v) = \frac{1}{\tau^{2/3}} \a_{1}(v) + \frac{1}{\tau^{4/3}} \a_{2}(v) + \frac{1}{\tau^{2}} \a_{3}(v)+\calo(\t^{-8/3})\\
&{\rm or}\\
&a(\tau,v) = \frac{1}{\tau^{1/3}} \hat{\a}_{1}(v) + \frac{1}{\tau} \hat{\a}_{2}(v) + \frac{1}{\tau^{5/3}} \hat{\a}_{3}(v)+\calo(\t^{-7/3})\,.\\
\end{split}
\eqlabel{ltseries3}
\end{equation}
For each of the four possible asymptotic expansions, we solve \eqref{Rtt}-\eqref{alphaeq} subject to the boundary
conditions
\begin{equation}
\begin{split}
&\bigg\{a_i(v),b_i(v),c_i(v)\bigg\}\bigg|_{v\to 0}=0\,,\qquad \bigg\{\phi_i(v)\ {\rm or}\ \hat{\phi}_i(v)\bigg\}
\bigg|_{v\to 0}=0\,,\\
&\bigg\{\a_i(v)\ {\rm or}\ \hat{\a}_i(v)\bigg\}
\bigg|_{v\to 0}=0\,.
\end{split}
\eqlabel{bbc}
\end{equation}
Additionally, we will require the absence of singularities in the asymptotic late-time expansion of the
quadratic curvature invariant $\cali^{[2]}$:
\begin{equation}
\begin{split}
\cali^{[2]}&\equiv \calr_{\mu\nu\r\l}\calr^{\mu\nu\r\l}\\
&=\cali^{[2]}_0(v)+\frac{1}{\t^{2/3}} \cali^{[2]}_1(v)+\frac{1}{\t^{4/3}} \cali^{[2]}_2(v)+\frac{1}{\t^{2}} \cali^{[2]}_3(v)+\calo(\t^{-8/3})\,.
\end{split}
\eqlabel{i2def}
\end{equation}
Of course, ultimately, all the curvature invariants  of the bulk geometry must be nonsingular for the supergravity approximation
dual to boost invariant $\caln=4$ SYM plasma to be consistent, unless singularities are hidden behind the horizon. However, it turns out that
asymptotic expansions  \eqref{ltseries1}-\eqref{ltseries3} are fixed unambiguously, given \eqref{bbc} and \eqref{i2def}. Thus
non-singularity of the other curvature invariants provides a strong consistency check on the validity of the supergravity approximation.

\subsection{Solution of the late-time series and curvature singularities}

The absence of curvature singularities in $\cali^{[2]}_0$ determines the leading solution  to be \cite{j1}
\begin{equation}
\begin{split}
a_0=\frac 12 \ln\frac{\left(1-v^4/3\right)^2}{1+v^4/3}\,,\qquad b_0=c_0=\frac 12 \ln\left(1+v^4/3\right)\,.
\end{split}
\eqlabel{order0}
\end{equation}
Notice that $v=3^{1/4}$ is the horizon to leading order in $\t$. It is straightforward to find that the  absence of naked singularities
in the bulk (specifically at $v=3^{1/4}$)  in $\cali^{[2]}_{1,2}$ would require that $\a_1=\a_2=0$\, $\hat{\a}_1=\hat{\a}_2=0$,
$\phi_1=\phi_2$ and $\hat{\phi}_1=\hat{\phi}_2=0$. We further have
\cite{shin1,j2,j3} 
\begin{equation}
\begin{split}
&a_1=\eta_0\ \frac{\left(9-v^4\right)v^4}{9-v^8}\,,\qquad  c_1=-\eta_0\ \frac{v^4}{3+v^4}-\frac{\eta_0}{2}\ \ln\frac{3-v^4}{3+v^4}\,, \\
&b_1=-3\eta_0\ \frac{v^4}{3+v^4}-2 c_1\,,
\end{split}
\eqlabel{order1}
\end{equation}
and
\begin{equation}
\begin{split}
a_2=&\frac{(9+5v^4)v^2}{12(9-v^8)} -C \frac{(9+v^4)v^4}{72(9-v^8)} +
\eta_0^2 \frac{(-1053-171v^4+9v^8+7v^{12})v^4}{6(9-v^8)^2}\\
&+
\frac{1}{8\sqrt{3}} \ln \frac{\sqrt{3}-v^2}{\sqrt{3}+v^2}-
\frac{3}{4} \eta_0^2 \ln \frac{3-v^4}{3+v^4}  \\
c_2 =& -\frac{\pi^2}{288\sqrt{3}} +\frac{v^2(9+v^4)}{12(9-v^8)} +C
\frac{v^4}{72(3+v^4)} -\eta_0^2 \frac{(-9+54v^4+7v^8)v^4}{6(3+v^4)(9-v^8)}\\
& +
\frac{1}{8\sqrt{3}} \ln \frac{\sqrt{3}-v^2}{\sqrt{3}+v^2}+\frac{1}{72} (C+66\eta_0^2) \ln \frac{3-v^4}{3+v^4}\\
&+
 \frac{1}{24\sqrt{3}} \left( \ln \frac{\sqrt{3}-v^2}{\sqrt{3}+v^2}
\ln\frac{(\sqrt{3}-v^2)(\sqrt{3}+v^2)^3}{4(3+v^4)^2} -{\rm li}_2 \left(-
\frac{(\sqrt{3}-v^2)^2}{(\sqrt{3}+v^2)^2} \right)\right)\\
b_2 =& -2 c_2+\frac{v^2}{4(3+v^4)} +C\frac{v^4}{24(3+v^4)} +\eta_0^2
\frac{(39+7v^4)v^4}{2(3+v^4)^2} +
\frac{1}{8\sqrt{3}} \ln \frac{\sqrt{3}-v^2}{\sqrt{3}+v^2}\\
&+
\frac{3}{4} \eta_0^2 \ln \frac{3-v^4}{3+v^4}\,,
\end{split}
\eqlabel{order2}
\end{equation}
with \cite{j2}
\begin{equation}
\eta_0=\frac{1}{2^{1/2}3^{3/4}}\,.
\eqlabel{eta0}
\end{equation}

While the equations at the third order for $\{a_3,b_3,c_3\}$ are a bit complicated, it is possible to decouple the equation
for $a_3$. The resulting equation is too long to be presented here. It is straightforward to solve the equation
perturbatively as
\begin{equation}
y\equiv 3-v^4\to 0_+ \,,
\eqlabel{ydef}
\end{equation}
\begin{equation}
\frac{d{a_3}}{dy}=-\frac{\sqrt{6}}{288}\ y^{-4}+\left( \frac{2^{1/2}3^{1/4}}{48}-\frac{2^{1/2}3^{1/4}\ln 2}{144}\right)\ y^{-3}+\calo\left(y^{-2}\right)\,,
\eqlabel{da3}
\end{equation}
which is sufficient to determine the singularities in $\cali^{[2]}_3$ at $v=3^{1/4}$.

The absence of pole singularities in $\cali^{[2]}_3$ at $v=3^{1/4}$ implies that only $\a_3\ne 0$:
\begin{equation}
\a_3=\a_{3,0}\left( \left(\frac{1}{96v^4}+\frac{v^4}{864}\right)\ \ln\frac{3+v^4}{3-v^4}-\frac{1}{144}\right)\,,
\eqlabel{a3solve}
\end{equation}
where $\a_{3,0}$ is an arbitrary constant; it further constraints \cite{j3}
\begin{equation}
C=2\sqrt{3}\ \ln 2-\frac{17}{\sqrt{3}}\,.
\eqlabel{cdef}
\end{equation}
Using \eqref{da3}-\eqref{cdef} we find
\begin{equation}
\cali^{[2]}_3={\rm finite}\ + \left(8\ 2^{1/2}\ 3^{3/4}+\frac{14}{3}\ \a_{3,0}\right)\ \ln(3-v^4)\,,\qquad v\to 3^{1/4}_-\,.
\eqlabel{lnsing}
\end{equation}
Notice that with $\a_{3,0}=0$ the logarithmic singularity agrees with the one found in \cite{j3}. From \eqref{lnsing}
it appears that the curvature singularities in the bulk of the supergravity dual to
expanding $\caln=4$ SYM plasma can be canceled for an appropriate choice of $\a_{3,0}$.
While the Ricci scalar $\calr$ is indeed nonsingular, unfortunately, this is not the case for the square of the
Ricci tensor.
We find
\begin{equation}
\calr_{\mu\nu}\calr^{\mu\nu}={\rm finite}+\frac{1}{\t^2}\ \frac{40}{3}\ \a_{3,0}\ \ln(3-v^4)\,,\qquad v\to 3^{1/4}_-\,.
\eqlabel{r2sing}
\end{equation}
Thus, it is impossible to cancel logarithmic singularity both in $\calr_{\mu\nu}\calr^{\mu\nu}$ and
$\calr_{\mu\nu\r\l}\calr^{\mu\nu\r\l}$ at order $\calo\left(\t^{-2}\right)$ with a supergravity field $\a$.

\subsection{Curvature singularities of the string frame metric}

In \cite{j3} it was suggested that the logarithmic singularity in the curvature invariants of the
Einstein frame metric \eqref{10dimM} might be canceled in string frame metric
\begin{equation}
g_{\mu\nu}^{string}\equiv e^{\phi/2}\ g_{\mu\nu}\,.
\eqlabel{sfme}
\end{equation}
Unfortunately, this is not possible. Indeed, as in \cite{j3}, in order to avoid pole singularities in the
curvature invariants of the string frame metric at leading and first three subleading orders,
the dilaton can contribute only at order $\calo(\t^{-2})$
\begin{equation}
\phi(\t,v)=\frac{1}{\t^{2}}\ k_3\ \ln\frac{3-v^4}{3+v^4}+\calo\left(\t^{-8/3}\right)\,,
\eqlabel{dilcon}
\end{equation}
where $k_3$ is an arbitrary constant.

We find then that
\begin{equation}
R_{\mu\nu}^{(10,string)}R^{(10,string)\mu\nu}={\rm finite}+\frac{1}{\t^2}\ \left(\frac{40}{3}\ \a_{3,0}-160\ k_3\right)\ \ln(3-v^4)\,,
\eqlabel{riccis}
\end{equation}
and\footnote{The result for the Riemann tensor squared here corrects the expression presented in \cite{j3}. }
\begin{equation}
R_{\mu\nu\r\l}^{(10,string)}R^{(10,string)\mu\nu\r\l}={\rm finite}+\frac{1}{\t^2}\ \left(8\ 2^{1/2}\ 3^{3/4}+\frac{14}{3}\ \a_{3,0}-152\ k_3\right)\ \ln(3-v^4)\,,
\eqlabel{riems}
\end{equation}
as $ v\to 3^{1/4}_-$.
While it is possible to remove logarithmic singularities in \eqref{riccis} and \eqref{riems} by properly adjusting
$\a_{3,0}$ and $k_3$, the higher curvature invariants of the string frame metric would remain necessarily singular.
Specifically, for string frame fourth order curvature invariants $\cali^{(10,string)[4]}_3$ we find
\begin{equation}
\begin{split}
\cali^{(10,string)[4]}_3\equiv& R^{(10,string)}_{\mu_1\nu_1\mu\nu}R^{(10,string)\mu_1\nu_1}\ _{\r\l}\ R^{(10,string)}_{\mu_2\nu_2}\ ^{\mu\nu}R^{(10,string)\mu_2\nu_2\r\l}
\\
=&{\rm finite}+\frac{1}{\t^2}\ \left(\frac{640}{3}\ 2^{1/2}\ 3^{3/4}+\frac{400}{3}\ \a_{3,0}-3904\ k_3\right)\ \ln(3-v^4)\,,
\end{split}
\eqlabel{i4string}
\end{equation}
as $ v\to 3^{1/4}_-$.

\section{KW QGP}

In the previous section we studied supergravity dual to $\caln=4$ SYM plasma assuming unbroken $SO(6)$ global symmetry
and the parity invariance along the extended boundary spacial directions.
At a technical level, we found that the absence of bulk singularities in the gravitational dual of expanding boost invariant plasma
is linked to nontrivial profiles of massive supergravity modes (corresponding to VEV's of irrelevant gauge invariant operators
in plasma). This suggests that turning on additional massive supergravity modes might remove curvature singularities.
Within the assumed unbroken symmetries of the $\caln=4$ plasma there are no additional massive modes in the supergravity approximation.
However, we can test the link between massive supergravity modes and singularities in the supergravity
dual of expanding boost invariant conformal plasmas for a slightly more  complicated example.

In this section we study supergravity dual to boost invariant expanding Klebanov-Witten superconformal plasma \cite{kw}.
Assuming that parity along extended boundary spacial directions as well as
the global $SU(2)\times SU(2)\times U(1)$ symmetry of KW plasma at equilibrium is unbroken, the effective supergravity description
will contain two massive modes \cite{aby1,bkt} --- a supergravity mode dual to a dimension-8 operator
( an analog of $\a$ field in the context of $\caln=4$ plasma ) and a supergravity mode dual to a dimension-6 operator.
We expect that appropriately exciting both modes we can remove curvature singularities in all quadratic invariants of the metric curvature
at the third order in late-time expansion. While we show that the latter expectation is correct, we also find that
higher curvature invariants will remain singular in this model. In fact, it appears we need an infinite set of massive fields (which is
not possible in the supergravity approximation)
to have a nonsingular metric.

Since the computations for the most part mimic the analysis of the previous section, we highlight only the
main results.

\subsection{Consistent Kaluza-Klein reduction}

Consistent KK reduction of the KW gauge theory plasma has been constructed in \cite{aby1,bkt}.

The five dimensional effective action is \cite{bkt}
\begin{equation}
\begin{split}
S_5= \frac{1}{16\pi G_5} \int_{\calm_5} {{\rm vol}}_{\calm_5}\
 \biggl\lbrace &
R_5-\frac{40}{3}(\del f)^2-20(\del w)^2-\frac 12 (\del\phi)^2-\calp
\biggr\rbrace\,,
\end{split}
\eqlabel{5actionE}
\end{equation}
where we defined
\begin{equation}
\calp=-24 e^{-\ft {16}{3}f-2w}+4 e^{-\ft {16}{3}f-12w}+\frac 12 K^2 e^{-\ft {40}{3} f }\,.
\eqlabel{calp}
\end{equation}
We set the asymptotic AdS radius to one, which corresponds to setting
\begin{equation}
K=4\,.
\eqlabel{setk}
\end{equation}

From Eq.~\eqref{5actionE} we obtain the following equations of motion
\begin{equation}
\begin{split}
0=\Box f-\frac{3}{80}\ \frac{\del\calp}{\del f}\,,
\end{split}
\eqlabel{eqf}
\end{equation}
\begin{equation}
\begin{split}
0=\Box w-\frac{1}{40}\ \frac{\del\calp}{\del w}\,,
\end{split}
\eqlabel{eqw}
\end{equation}
\begin{equation}
\begin{split}
0=\Box \phi- \frac{\del\calp}{\del \phi}\,,
\end{split}
\eqlabel{eqphi}
\end{equation}
\begin{equation}
\begin{split}
R_{5\mu\nu}=&\frac{40}{3}\ \del_\mu f\del_\nu f+20\ \del_\mu w\del_\nu w+\frac 12\  \del_\mu \phi\del_\nu \phi+\frac 13 g_{\mu\nu}\ \calp\,.
\end{split}
\eqlabel{eqE}
\end{equation}

The uplifted ten dimensional metric takes form
\begin{equation}
ds_{10}^2 =g_{\mu\nu}(y) dy^{\mu}dy^{\nu}+\om_1^2(y) e_{\psi}^2
+\om_2^2(y) \sum_{a=1}^2\left(e_{\theta_a}^2+e_{\phi_a}^2\right)\,,
\eqlabel{10met}
\end{equation}
where $y$ denotes the coordinates of $\calm_5$ (Greek indices $\mu,\nu$
will run from $0$ to $4$) and
the one-forms $e_{\psi},\ e_{\theta_a},\ e_{\phi_a}$ ($a=1,2$) are given by
\begin{equation}
e_{\psi}=\frac 13 \left(d\psi+\sum_{a=1}^2 \cos\theta_a\
d\phi_a\right)\,,\qquad
e_{\theta_a}=\frac{1}{\sqrt{6}} d\theta_a\,,\qquad
e_{\phi_a}=\frac{1}{\sqrt{6}} \sin\theta_a\ d\phi_a\,.
\eqlabel{1forms}
\end{equation}
Furthermore,
\begin{equation}
g_{\mu\nu}(y) dy^{\mu}dy^{\nu}=\om_1^{-2/3}\om_2^{-8/3}\ ds^2\,,\qquad \om_1=e^{f-4w}\,,\qquad \om_2=e^{f+w}\,,
\eqlabel{5drel}
\end{equation}
and $ds^2$ is the five-dimensional metric \eqref{5dimM}.

\subsection{Late-time expansion and solution}

We consider the same five-dimensional metric ansatz as in \eqref{5dimM}; we use the
late-$\t$ expansion of the metric warp factors $\{a,b,c\}$ as in \eqref{ltseries1}.
In order to avoid pole singularities in  curvature invariants, the dilaton must be set to zero.
For the supergravity scalar $f(\t,v)$ dual to a dimension-8 operator in KW plasma and for
the supergravity scalar $w(\t,v)$ dual to a dimension-6 operator in KW plasma
we use the asymptotics
\begin{equation}
\begin{split}
&f(\tau,v) =\frac{1}{\tau^{2}} f_3(v)+\calo\left(\t^{-8/3}\right)\\
&w(\tau,v) =\frac{1}{\tau^{2}} w_3(v)+\calo\left(\t^{-8/3}\right)\,.
\end{split}
\eqlabel{ltseries4}
\end{equation}
Since the massive supergravity modes $\{f,w\}$ are turned on only at order $\calo(\t^{-2})$,
the metric warp factors $\{a_i(v),b_i(v),c_i(v)\}$ are exactly the same as for the $\caln=4$
SYM plasma, see \eqref{order0}-\eqref{order2}, \eqref{eta0}, \eqref{da3} and  \eqref{cdef}.
Moreover, analogously to \eqref{a3solve}
\begin{equation}
f_3(v)=f_{3,0}\left( \left(\frac{1}{96v^4}+\frac{v^4}{864}\right)\ \ln\frac{3+v^4}{3-v^4}-\frac{1}{144}\right)\,.
\eqlabel{f3solve}
\end{equation}
For $w_3$ we find the following equation
\begin{equation}
0=w_3''+\frac{5 v^8+27}{v(v^8-9)}\ w_3'-\frac{12}{v^2}\ w_3\,.
\eqlabel{w3}
\end{equation}
Solution to $w_3$ must have a vanishing non-normalizable mode as $v\to 0$.
Near the horizon \eqref{ydef}, the most general solution to \eqref{w3} takes form
\begin{equation}
w_3=w_{0,0}+w_{0,1}\ln\left(4\ 3^{3/4}\ y\right)+\calo\left(y\right)\,.
\eqlabel{w3h}
\end{equation}
We find  that vanishing of the non-normalizable mode of $w_3$ as $v\to 0$
requires
\begin{equation}
w_{0,0}=w_{0,1}\ \times\ \left[2 \gamma - \ln\left({\frac{3}{2}}\right) + \psi\left(\frac{1}{2}\right) + \psi\left(\frac{3}{2}\right)\right]\,,
\eqlabel{w00}
\end{equation}
in which case,
\begin{equation}
w_3= -\frac{\pi}{6 \sqrt{3}} \times\ w_{0,1}\ v^{6}+\calo\left(v^{14}\right)\,.
\eqlabel{boun}
\end{equation}

\subsection{Quadratic curvature invariants of \eqref{10met}}

We collect here results for $\calr$, $\calr_{\mu\nu}\calr^{\mu\nu}$ and $\calr_{\mu\nu\r\l}\calr^{\mu\nu\r\l}$
curvature invariants of the metric \eqref{10met} to the third order in the late-$\t$ expansion.

\subsubsection{At leading order}

\begin{equation}
\calr\bigg|^{(0)}=-20+20\,,
\eqlabel{rs0}
\end{equation}
\begin{equation}
\calr_{\mu\nu}\calr^{\mu\nu}\bigg|^{(0)}=80+80\,,
\eqlabel{ri0}
\end{equation}
\begin{equation}
\calr_{\mu\nu\r\l}\calr^{\mu\nu\r\l}\bigg|^{(0)}=\frac{8 (5 v^{16}+60v^{12}+1566v^8+540v^4+405)}{(3+v^4)^4}+136\,,
\eqlabel{re0}
\end{equation}
where in \eqref{rs0}-\eqref{re0} we separated the $AdS_5$ and the $T^{1,1}$ contributions.

\subsubsection{At first order}

\begin{equation}
\calr\bigg|^{(1)}=0\,,
\eqlabel{rs1}
\end{equation}
\begin{equation}
\calr_{\mu\nu}\calr^{\mu\nu}\bigg|^{(1)}=0\,,
\eqlabel{ri1}
\end{equation}
\begin{equation}
\calr_{\mu\nu\r\l}\calr^{\mu\nu\r\l}\bigg|^{(1)}=\frac{1}{\t^{2/3}}\ \frac{41472 (v^4-3)v^8}{(3+v^4)^5}\ \eta_0\,.
\eqlabel{re1}
\end{equation}

\subsubsection{At second order}

\begin{equation}
\calr\bigg|^{(2)}=0\,,
\eqlabel{rs2}
\end{equation}
\begin{equation}
\calr_{\mu\nu}\calr^{\mu\nu}\bigg|^{(2)}=0\,,
\eqlabel{ri2}
\end{equation}
\begin{equation}
\begin{split}
&\calr_{\mu\nu\r\l}\calr^{\mu\nu\r\l}\bigg|^{(2)}=\frac{1}{\t^{2/3}}\ \biggl\{
- \frac {576 (v^4-3)v^8}{(3+v^4)^5} C\\
&+\frac{6912(5v^{24}-60v^{20} +2313
v^{16}  -6912 v^{12}+26487 v^8-18468 v^4+13851)v^8 \eta_0^2}{(3-v^4)^4
(3+v^4)^6}\\
&-\frac{4608(5 v^{16}+6 v^{12}+162 v^8 +54 v^4+405) v^{10}}{(3-v^4)^4
  (3+v^4)^5} \biggr\}\,.
\end{split}
\eqlabel{re2}
\end{equation}

\subsubsection{At third order}

Provided that $C$ is chosen as in \eqref{cdef}, we  find that
\begin{equation}
\calr\bigg|^{(3)}=0\,,
\eqlabel{rs3}
\end{equation}
once we use Einstein equations.
Also, it is easy to determine that
\begin{equation}
\calr_{\mu\nu}\calr^{\mu\nu}\bigg|^{(3)}=-f_{3,0}\ \times\  \frac{100}{27}\left(\left(v^4+\frac{9}{v^4}\right)\ln\frac{3+v^4}{3-v^4}-6\right)\,.
\eqlabel{ri3}
\end{equation}
The non-singularity condition therefore requires that
\begin{equation}
f_{3\,0}=0\,.
\eqlabel{f30}
\end{equation}

Finally, using \eqref{da3}, \eqref{w3h} and \eqref{f30}, we find
\begin{equation}
\begin{split}
\calr_{\mu\nu\r\l}\calr^{\mu\nu\r\l}\bigg|^{(3)}=&\left(8\ 2^{1/2}\ 3^{3/4}-384 w_{0,1}\right)\ \ln(3-v^4)\\
&-\left(\frac{20}{3}\ 2^{1/2}\ 3^{3/4}\ \ln\left(6\ (3e)^{1/5}\right)+384 w_{0,0}\right)+\calo\left((3-v^4)\right)\,.
\end{split}
\eqlabel{re3}
\end{equation}
Thus,  choosing
\begin{equation}
w_{0,1}=\frac{2^{1/2}\ 3^{3/4}}{48}\,,
\eqlabel{w30}
\end{equation}
the logarithmic singularity in \eqref{re3} is removed.

\subsection{Higher order curvature invariants of \eqref{10met}}

Let us define shorthand notation for the contractions of the Riemann tensor. For each integer $n$ we have
\begin{equation}
\calr^{[2^n]}\ _{\mu\nu\r\l}\equiv \calr^{[2^{n-1}]}\ _{\mu_1\nu_1\mu\nu}\cdot \calr^{[2^{n-1}]}\ ^{\mu_1\nu_1}\ _{\r\l}\,,
\eqlabel{riemann}
\end{equation}
where
\begin{equation}
\calr^{[0]}\ _{\mu\nu\r\l}\equiv \calr_{\mu\nu\r\l}\,.
\eqlabel{n0}
\end{equation}
We further define higher curvature invariants $\cali^{[2^n]}$, generalizing \eqref{i2def}:
\begin{equation}
\begin{split}
\cali^{[2^n]}&\equiv \calr^{[2^{n-1}]}\ _{\mu\nu\r\l}\calr^{[2^{n-1}]}\ ^{\mu\nu\r\l}\\
&=\cali^{[2^n]}_0(v)+\frac{1}{\t^{2/3}} \cali^{[2^n]}_1(v)+\frac{1}{\t^{4/3}} \cali^{[2^n]}_2(v)+\frac{1}{\t^{2}} \cali^{[2^n]}_3(v)+\calo(\t^{-8/3})\,.
\end{split}
\eqlabel{indef}
\end{equation}
With a straightforward albeit tedious computation we can extract logarithmic singularities in $\cali^{[2^n]}_3$.
For the first couple invariants we find:
\begin{equation}
\begin{split}
\cali^{[2]}_3=&-384\, \left( w_{0,1}-\frac{1}{48}\,{3}^{3/4}\sqrt {2} \right) \ln  \left( 3
-{v}^{4} \right)+{\rm finite}\\
\cali^{[4]}_3=&-3072\, \left( w_{0,1}-{\frac {5}{72}}\,{3}^{3/4}\sqrt {2} \right)
\ln  \left( 3-{v}^{4} \right)
+{\rm finite}\\
\cali^{[8]}_3=&-98304\, \left( w_{0,1}-{\frac {14}{27}}\,{3}^{3/4}\sqrt {2}
 \right) \ln  \left( 3-{v}^{4} \right)
+{\rm finite}\\
\cali^{[16]}_3=&-50331648\, \left( w_{0,1}+{\frac {512}{3}}\,{3}^{3/4}\sqrt {2}
 \right) \ln  \left( 3-{v}^{4} \right)
+{\rm finite}\\
\cali^{[32]}_3=&-6597069766656\, \left( w_{0,1}+{\frac {1375731712}{27}}\,{3}^{3/4}
\sqrt {2} \right) \ln  \left( 3-{v}^{4} \right)
+{\rm finite}\\
\cali^{[64]}_3=&-56668397794435742564352\, \left( w_{0,1}+{\frac {
5044031582654955520}{9}}\,{3}^{3/4}\sqrt {2} \right)\\
&\times \ln  \left( 3-{v}
^{4} \right)+{\rm finite}\,,\\
\end{split}
\eqlabel{higher}
\end{equation}
as $v\to 3^{1/4}_-$.
Interestingly, we find that all lower order invariants, \ie,
\begin{equation}
\cali^{[2^n]}_{i}\,,\qquad n=\{1,2,3,4,5,6\}\,,\qquad i=\{0,1,2\}\,,
\eqlabel{higherfinite}
\end{equation}
are finite as $v\to 3^{1/4}_-$.

Clearly, given \eqref{higher}, logarithmic singularities of the curvature invariant \eqref{indef} can not be canceled within the
supergravity approximation.

\section{Conclusion}

In this paper, following \cite{j1,j2,j3} we attempted to construct a string theory  dual to strongly coupled conformal expanding
plasmas in Bjorken regime \cite{bj}. In order to have computational control we truncated the full string theory to supergravity
approximation, and focused on the well-established examples of the gauge/string dualities:
we considered $\caln=4$ SYM \cite{m1} and superconformal Klebanov-Witten \cite{kw} gauge theories. We used
non-singularity of the dual gravitational backgrounds as a guiding principle to identify gauge theory
operators that would develop a vacuum expectation value during boost-invariant expansion of the plasma. Truncation to
a supergravity sector of the string theory (along with parity invariance in the Bjorken frame) severely restricts a
set of such operators. In the case of the $\caln=4$  SYM, there are only two such gauge invariant operators, while
for the Klebanov-Witten plasma one has an additional operator. We constructed supergravity dual as a late-time
asymptotic expansion and demonstrated that the gravitational boundary stress energy tensor expectation value has exactly the
same asymptotic late-time expansion as predicted by M\"uller-Israel-Stewart theory of transient relativistic kinetic  theory \cite{m,is} for the
boost invariant expansion. As an impressive success of this approach, one recovers by requiring
non-singularity of the background geometry at leading and first three subleading orders\footnote{ At the third
subleading order, logarithmic singularities of the metric curvature invariants remain.} \cite{j1,j2,j3} the
equation of state for the plasma, its shear viscosity and its relaxation time, in agreement with values extracted from the
equilibrium higher point correlation functions. Unfortunately, we showed that logarithmic singularity in the
background geometry can not be canceled within the supergravity approximation. Moreover, given that the singularities
appear to persist in arbitrary high metric curvature invariants, we suspect that relaxing the constraint of parity invariance in the
Bjorken regime would not help. Indeed, relaxing parity invariance would allow for only finite number of additional
(massive) supergravity modes, which, as an example of Klebanov-Witten plasma demonstrates, would only allow to cancel
logarithmic singularities in finite number of additional metric curvature invariants.

We would like to conclude with several speculations.
\nxt It is possible that though the full asymptotic late-time expansion of strongly coupled expanding boost invariant plasma
is ill-defined within supergravity approximation, the first
couple orders can nonetheless we used to extract transport coefficients, and thus can be of value to RHIC (and
future LHC) experiments. To this end we note a curiosity that the relaxation time of the $\caln=4$ and the KW plasma
came up to be the same. Moreover turning on the scalar fields
did not modify the extracted relaxation time.
This might point to the universality of the relaxation time, not too dissimilar
to the universality of the shear viscosity in gauge theory plasma at (infinitely) strong 't Hooft coupling (see 
 \cite{urelax} for a conjecture on a bound on relaxation times).
Second, the relaxation time computed is the relaxation time for the equilibration of the shear modes.
If there is substantial bulk viscosity at RHIC, one would also need an estimate for the corresponding relaxation time ---
this is where non-conformal gauge/string dualities might be useful.
\nxt It is possible that the singularity observed in the supergravity approximation
is genuine, and can not be cured by string theory corrections. A prototypical example of this is the
Klebanov-Tseytlin solution \cite{kt}. There, one does not expect a resolution of the singularity by string corrections
(while preserving the chiral symmetry and supersymmetry) simply because the corresponding (dual) gauge theory phase does not
exist. By the same token, the observed singularity in the string theory dual to strongly coupled boost invariant expanding
plasma might indicate that such a flow for a {\it conformal} plasma is physically impossible to realize. For example,
a Bjorken flow of a conformal plasma might always  become turbulent at
late times.  In fact, at weak coupling, rapidity breaking
instabilities were found \cite{instab,instab2}. Although this was far from a
hydrodynamical regime, the leftover logarithmic singularity appearing
at the third order might be a manifestation of such an effect.

We hope to address these issues in future work.

\section*{Acknowledgments}
We would like to thank  Xiao Liu, Rob Myers, Andrei Starinets and  Sam Vazquez for valuable discussion.
AB's research at Perimeter Institute is supported in part by the Government
of Canada through NSERC and by the Province of Ontario through MRI.
AB gratefully acknowledges further support by an NSERC Discovery
grant and support through the Early Researcher Award program by the
Province of Ontario. RJ and MH were supported in part by Polish
Ministry of Science and Information Technologies grant 1P03B04029
(2005-2008), RTN network ENRAGE MRTN-CT-2004-005616, and the Marie
Curie ToK COCOS (contract MTKD-CT-2004-517186).


\begin{thebibliography}{99}

\bibitem{m1}
  J.~M.~Maldacena,
  ``The large N limit of superconformal field theories and supergravity,''
  Adv.\ Theor.\ Math.\ Phys.\  {\bf 2}, 231 (1998)
  [Int.\ J.\ Theor.\ Phys.\  {\bf 38}, 1113 (1999)]
  [arXiv:hep-th/9711200].

\bibitem{m2}
  O.~Aharony, S.~S.~Gubser, J.~M.~Maldacena, H.~Ooguri and Y.~Oz,
  ``Large N field theories, string theory and gravity,''
  Phys.\ Rept.\  {\bf 323}, 183 (2000)
  [arXiv:hep-th/9905111].


\bibitem{rhic1}
  K.~Adcox {\it et al.}  [PHENIX Collaboration],
  ``Formation of dense partonic matter in relativistic nucleus nucleus
  collisions at RHIC: Experimental evaluation by the PHENIX  collaboration,''
  Nucl.\ Phys.\ A {\bf 757}, 184 (2005)
  [arXiv:nucl-ex/0410003].

\bibitem{rhic2}
  B.~B.~Back {\it et al.},
  ``The PHOBOS perspective on discoveries at RHIC,''
  Nucl.\ Phys.\ A {\bf 757}, 28 (2005)
  [arXiv:nucl-ex/0410022].

\bibitem{rhic3}
  I.~Arsene {\it et al.}  [BRAHMS Collaboration],
  ``Quark gluon plasma and color glass condensate at RHIC? The perspective
  from the BRAHMS experiment,''
  Nucl.\ Phys.\ A {\bf 757}, 1 (2005)
  [arXiv:nucl-ex/0410020].

\bibitem{rhic4}
  J.~Adams {\it et al.}  [STAR Collaboration],
  ``Experimental and theoretical challenges in the search for the quark  gluon
  plasma: The STAR collaboration's critical assessment of the  evidence from
  RHIC collisions,''
  Nucl.\ Phys.\ A {\bf 757}, 102 (2005)
  [arXiv:nucl-ex/0501009].

\bibitem{ads1}
  S.~S.~Gubser, I.~R.~Klebanov and A.~W.~Peet,
  Phys.\ Rev.\ D {\bf 54}, 3915 (1996)
  [arXiv:hep-th/9602135].

\bibitem{ads2}
  S.~S.~Gubser, I.~R.~Klebanov and A.~A.~Tseytlin,
  ``Coupling constant dependence in the thermodynamics of N = 4  supersymmetric
  Yang-Mills theory,''
  Nucl.\ Phys.\ B {\bf 534}, 202 (1998)
  [arXiv:hep-th/9805156].


\bibitem{nc1}
  A.~Buchel, S.~Deakin, P.~Kerner and J.~T.~Liu,
  ``Thermodynamics of the N = 2* strongly coupled plasma,''
  Nucl.\ Phys.\  B {\bf 784}, 72 (2007)
  [arXiv:hep-th/0701142].

\bibitem{nc2}
  O.~Aharony, A.~Buchel and P.~Kerner,
  ``The black hole in the throat - thermodynamics of strongly coupled
  cascading gauge theories,''
  Phys.\ Rev.\  D {\bf 76}, 086005 (2007)
  [arXiv:0706.1768 [hep-th]].


\bibitem{nc3}
  D.~Mateos, R.~C.~Myers and R.~M.~Thomson,
  ``Thermodynamics of the brane,''
  JHEP {\bf 0705}, 067 (2007)
  [arXiv:hep-th/0701132].


\bibitem{u1}
  A.~Buchel and J.~T.~Liu,
  ``Universality of the shear viscosity in supergravity,''
  Phys.\ Rev.\ Lett.\  {\bf 93}, 090602 (2004)
  [arXiv:hep-th/0311175].

\bibitem{u2}
  P.~Kovtun, D.~T.~Son and A.~O.~Starinets,
  ``Viscosity in strongly interacting quantum field theories from black hole
  physics,''
  Phys.\ Rev.\ Lett.\  {\bf 94}, 111601 (2005)
  [arXiv:hep-th/0405231].

\bibitem{u3}
  A.~Buchel,
  ``On universality of stress-energy tensor correlation functions in
  supergravity,''
  Phys.\ Lett.\  B {\bf 609}, 392 (2005)
  [arXiv:hep-th/0408095].

\bibitem{bulk}
  A.~Buchel,
  ``Bulk viscosity of gauge theory plasma at strong coupling,''
  arXiv:0708.3459 [hep-th].



\bibitem{cor1}
  A.~Buchel, J.~T.~Liu and A.~O.~Starinets,
  ``Coupling constant dependence of the shear viscosity in N=4 supersymmetric
  Yang-Mills theory,''
  Nucl.\ Phys.\  B {\bf 707}, 56 (2005)
  [arXiv:hep-th/0406264].

\bibitem{cor2}
  P.~Benincasa and A.~Buchel,
  ``Transport properties of N = 4 supersymmetric Yang-Mills theory at  finite
  coupling,''
  JHEP {\bf 0601}, 103 (2006)
  [arXiv:hep-th/0510041].


\bibitem{mt}
  J.~Mas and J.~Tarrio,
  ``Hydrodynamics from the Dp-brane,''
  JHEP {\bf 0705}, 036 (2007)
  [arXiv:hep-th/0703093].


\bibitem{lrw}
  H.~Liu, K.~Rajagopal and U.~A.~Wiedemann,
  ``Calculating the jet quenching parameter from AdS/CFT,''
  Phys.\ Rev.\ Lett.\  {\bf 97}, 182301 (2006)
  [arXiv:hep-ph/0605178].


\bibitem{hkkky}
  C.~P.~Herzog, A.~Karch, P.~Kovtun, C.~Kozcaz and L.~G.~Yaffe,
  ``Energy loss of a heavy quark moving through N = 4 supersymmetric
  Yang-Mills plasma,''
  JHEP {\bf 0607}, 013 (2006)
  [arXiv:hep-th/0605158].


\bibitem{aev}
  P.~C.~Argyres, M.~Edalati and J.~F.~Vazquez-Poritz,
  ``Spacelike strings and jet quenching from a Wilson loop,''
  JHEP {\bf 0704}, 049 (2007)
  [arXiv:hep-th/0612157].

\bibitem{CaronHuot:2006te}
  S.~Caron-Huot, P.~Kovtun, G.~D.~Moore, A.~Starinets and L.~G.~Yaffe,
  JHEP {\bf 0612}, 015 (2006)
  [arXiv:hep-th/0607237].


\bibitem{sw1}
  S.~S.~Gubser, S.~S.~Pufu and A.~Yarom,
  ``Sonic booms and diffusion wakes generated by a heavy quark in thermal
  AdS/CFT,''
  arXiv:0706.4307 [hep-th].

\bibitem{sw2}
  P.~M.~Chesler and L.~G.~Yaffe,
  ``The wake of a quark moving through a strongly-coupled $\mathcal N=4$
  supersymmetric Yang-Mills plasma,''
  Phys.\ Rev.\ Lett.\  {\bf 99}, 152001 (2007)
  [arXiv:0706.0368 [hep-th]].


\bibitem{j1}
  R.~A.~Janik and R.~Peschanski,
  ``Asymptotic perfect fluid dynamics as a consequence of AdS/CFT,''
  Phys.\ Rev.\  D {\bf 73}, 045013 (2006)
  [arXiv:hep-th/0512162].

\bibitem{shin1}
  S.~Nakamura and S.~J.~Sin,
  ``A holographic dual of hydrodynamics,''
  JHEP {\bf 0609} (2006) 020
  [arXiv:hep-th/0607123].

\bibitem{j2}
  R.~A.~Janik,
  ``Viscous plasma evolution from gravity using AdS/CFT,''
  Phys.\ Rev.\ Lett.\  {\bf 98}, 022302 (2007)
  [arXiv:hep-th/0610144].


\bibitem{j3}
   M.~P.~Heller and R.~A.~Janik,
     ``Viscous hydrodynamics relaxation time from AdS/CFT,''
       Phys.\ Rev.\  D {\bf 76} (2007) 025027
         [arXiv:hep-th/0703243].

\bibitem{zahed}
  E.~Shuryak, S.~J.~Sin and I.~Zahed,
  ``A gravity dual of RHIC collisions,''
  J.\ Korean Phys.\ Soc.\  {\bf 50}, 384 (2007)
  [arXiv:hep-th/0511199].

\bibitem{siopsis}
  J.~Alsup, C.~Middleton and G.~Siopsis,
  ``AdS/CFT Correspondence with Heat Conduction,''
  Phys.\ Lett.\  B {\bf 654}, 35 (2007)
  [arXiv:hep-th/0607139].

\bibitem{ext0}
  R.~A.~Janik and R.~Peschanski,
  ``Gauge / gravity duality and thermalization of a boost-invariant perfect
  fluid,''
  Phys.\ Rev.\  D {\bf 74}, 046007 (2006)
  [arXiv:hep-th/0606149].

\bibitem{shin2}
  S.~J.~Sin, S.~Nakamura and S.~P.~Kim,
  ``Elliptic flow, Kasner universe and holographic dual of RHIC fireball,''
  JHEP {\bf 0612}, 075 (2006)
  [arXiv:hep-th/0610113].

\bibitem{ext1}
  D.~Bak and R.~A.~Janik,
  ``From static to evolving geometries: R-charged hydrodynamics from
  supergravity,''
  Phys.\ Lett.\  B {\bf 645}, 303 (2007)
  [arXiv:hep-th/0611304].

\bibitem{kaj1}
  K.~Kajantie and T.~Tahkokallio,
  ``Spherically expanding matter in AdS/CFT,''
  Phys.\ Rev.\  D {\bf 75}, 066003 (2007)
  [arXiv:hep-th/0612226].

\bibitem{kov}
  Y.~V.~Kovchegov and A.~Taliotis,
  ``Early time dynamics in heavy ion collisions from AdS/CFT correspondence,''
  Phys.\ Rev.\  C {\bf 76}, 014905 (2007)
  [arXiv:0705.1234 [hep-ph]].

\bibitem{kaj2}
  K.~Kajantie, J.~Louko and T.~Tahkokallio,
  ``Gravity dual of 1+1 dimensional Bjorken expansion,''
  Phys.\ Rev.\  D {\bf 76}, 106006 (2007)
  [arXiv:0705.1791 [hep-th]].

\bibitem{ext2}
  J.~Grosse, R.~A.~Janik and P.~Surowka,
  ``Flavors in an expanding plasma,''
  arXiv:0709.3910 [hep-th].


\bibitem{bj}
  J.~D.~Bjorken,
  ``Highly Relativistic Nucleus-Nucleus Collisions: The Central Rapidity
  Region,''
  Phys.\ Rev.\  D {\bf 27}, 140 (1983).


\bibitem{pss}
  G.~Policastro, D.~T.~Son and A.~O.~Starinets,
  ``The shear viscosity of strongly coupled N = 4 supersymmetric Yang-Mills
  plasma,''
  Phys.\ Rev.\ Lett.\  {\bf 87}, 081601 (2001)
  [arXiv:hep-th/0104066].


\bibitem{st} The relaxation time of the $\caln=4$ plasma can also be extracted from equilibrium
correlation functions of the stress-energy tensor, A.~Starinets, private communication.
Also: S.~Vazquez, ``Generalized Hydrodynamics for
Strongly Coupled Plasmas'', to appear.


\bibitem{rrs}
  A.~Rebhan, P.~Romatschke and M.~Strickland,
  ``Hard-loop dynamics of non-Abelian plasma instabilities,''
  Phys.\ Rev.\ Lett.\  {\bf 94}, 102303 (2005)
  [arXiv:hep-ph/0412016].



\bibitem{kl}
  C.~P.~Herzog, I.~R.~Klebanov and P.~Ouyang,
  ``D-branes on the conifold and N = 1 gauge / gravity dualities,''
  arXiv:hep-th/0205100.


\bibitem{kt1}
  A.~Buchel,
  ``Finite temperature resolution of the Klebanov-Tseytlin singularity,''
  Nucl.\ Phys.\  B {\bf 600}, 219 (2001)
  [arXiv:hep-th/0011146].

\bibitem{kt2}
  A.~Buchel, C.~P.~Herzog, I.~R.~Klebanov, L.~A.~Pando Zayas and A.~A.~Tseytlin,
  ``Non-extremal gravity duals for fractional D3-branes on the conifold,''
  JHEP {\bf 0104}, 033 (2001)
  [arXiv:hep-th/0102105].


\bibitem{aby1}
  O.~Aharony, A.~Buchel and A.~Yarom,
  ``Holographic renormalization of cascading gauge theories,''
  Phys.\ Rev.\  D {\bf 72}, 066003 (2005)
  [arXiv:hep-th/0506002].


\bibitem{kt3}
  S.~S.~Gubser, C.~P.~Herzog, I.~R.~Klebanov and A.~A.~Tseytlin,
  ``Restoration of chiral symmetry: A supergravity perspective,''
  JHEP {\bf 0105}, 028 (2001)
  [arXiv:hep-th/0102172].


\bibitem{m} I.~M\"uller, Z.\ Phys. {\bf 198} (1967) 329.

\bibitem{is}
  W.~Israel and J.~M.~Stewart,
  ``Transient relativistic thermodynamics and kinetic theory,''
  Annals Phys.\  {\bf 118} (1979) 341.


\bibitem{mur}
  A.~Muronga,
  ``Causal Theories of Dissipative Relativistic Fluid Dynamics for Nuclear
  Collisions,''
  Phys.\ Rev.\  C {\bf 69}, 034903 (2004)
  [arXiv:nucl-th/0309055].



\bibitem{bkt}
  A.~Buchel,
  ``Transport properties of cascading gauge theories,''
  Phys.\ Rev.\  D {\bf 72}, 106002 (2005)
  [arXiv:hep-th/0509083].


\bibitem{kw}
  I.~R.~Klebanov and E.~Witten,
  ``Superconformal field theory on threebranes at a Calabi-Yau  singularity,''
  Nucl.\ Phys.\  B {\bf 536}, 199 (1998)
  [arXiv:hep-th/9807080].



\bibitem{BRW}
  R.~Baier, P.~Romatschke and U.~A.~Wiedemann,
  ``Dissipative hydrodynamics and heavy ion collisions,''
  Phys.\ Rev.\  C {\bf 73}, 064903 (2006)
  [arXiv:hep-ph/0602249].

\bibitem{urelax}
  S.~Hod,
  ``Universal bound on dynamical relaxation times and black-hole  quasinormal
  ringing,''
  Phys.\ Rev.\  D {\bf 75}, 064013 (2007)
  [arXiv:gr-qc/0611004].


\bibitem{kt}
  I.~R.~Klebanov and A.~A.~Tseytlin,
  ``Gravity duals of supersymmetric SU(N) x SU(N+M) gauge theories,''
  Nucl.\ Phys.\  B {\bf 578}, 123 (2000)
  [arXiv:hep-th/0002159].

\bibitem{instab}
  P.~Romatschke and R.~Venugopalan,
  ``Collective non-Abelian instabilities in a melting color glass
  condensate,''
  Phys.\ Rev.\ Lett.\  {\bf 96}, 062302 (2006)
  [arXiv:hep-ph/0510121].

\bibitem{instab2}
  P.~Romatschke and R.~Venugopalan,
  ``The unstable Glasma,''
  Phys.\ Rev.\  D {\bf 74}, 045011 (2006)
  [arXiv:hep-ph/0605045].

\end{thebibliography}
\end{document}